\documentclass[a4paper]{jpconf}
\usepackage{amsmath}
\usepackage{graphicx}
\begin{document}
\nocite{*}
%\title{Phenomenology of axion-photon coupling in the jets of AGNs}
\title{Phenomenology of axion-like particles coupling with photons in the jets of active galactic nuclei}
\author{Ahmed Ayad and Geoff Beck}
\address{School of Physics, University of the Witwatersrand, Private Bag 3, WITS-2050, Johannesburg, South Africa}
\ead{ahmed@aims.edu.gh}
%---------------------------------------------------------------------------------------
%---------------------------------------------------------------------------------------
\begin{abstract}
Axions or more generally axion-like particles (ALPs) are pseudo-scalar particles predicted by many extensions of the Standard Model of particle physics (SM) and considered as highly viable candidates for dark matter (DM) in the universe. If they exist in nature, they are expected to couple with photons in the presence of an external magnetic field through a form of the Primakoff effect. In this work, we examine the detectability of signals produced by ALP-photon coupling in the highly magnetized environment of the relativistic jets produced by active galactic nuclei (AGNs). Furthermore, we test a cosmic background ALP model, motivated by its explanation of the Coma cluster soft X-ray excess, in the environment of the M87 AGN jet. We then demonstrate the potential of the AGN jet environment to probe low-mass ALP models and to potentially constrain the model proposed to explain the Coma cluster X-ray excess.
\end{abstract}

\section{Introduction}
An outstanding result of modern cosmology is that only a small fraction of the total matter content of the universe is made of baryonic matter, while the vast majority is constituted by dark matter (DM) \cite{komatsu2011seven}. However, the nature of such a component is still unknown and might be a matter of long-standing controversies. In principle, the nature of DM can be understood through looking for light scalar candidates of DM such as axions and axion-like particles (ALPs). Axions \cite{peccei1977cp, weinberg1978new} are pseudo-Nambu-Goldstone bosons that appears after the spontaneous breaking of the Peccei-Quinn symmetry and it was introduced to solve the CP-violation problem2 of the strong interactions, which represents one of the serious problems in the standard model of particle physics (SM), for a review see reference \cite{peccei2008strong}. The theory, together with observational and experimental bounds, predicts that such axions are very light and weakly interacting with the SM particles, see Ref. \cite{asztalos2006searches}.  For these reasons, they are suggested to be suitable candidates for the DM content of the universe \cite{preskill1983cosmology, abbott1983cosmological, dine1983not}. The observation of a light axion would indeed solve the strong CP-problem and at least would participate in improving our understanding of the origin of the component of the DM in the universe. Furthermore, there are plenty of theories beyond the SM predict the existence of many other pseudo-scalar particles sharing the same phenomenology of the QCD axions \cite{arvanitaki2010string, cicoli2012type, anselm1982second}. Therefore, we strongly believe that axions or now more generally ALPs \cite{arias2012wispy, ringwald2012exploring, marsh2016axion} are a highly viable candidate for DM in the universe.

If they really exist in nature, ALPs are expected to couple with photons in the presence of an external electromagnetic field through the Primakoff effect \cite{sikivie1983experimental}. This coupling gives rise to the mixing of photons with ALPs \cite{raffelt1988mixing}, which leads to the conversion between photons and ALPs and changes the polarization state of photons. Over the last few years, it has been realized that this phenomenon would allow searches for the ALPs in the observations of distant AGNs in radio galaxies \cite{bassan2010axion, horns2012probing}. Since photons emitted by these sources can mix with ALPs during their propagation in the presence of an external magnetic field and this might reduce photon absorption caused by extragalactic background light \cite{harris2014photon}. Recent observations of blazars by the Fermi Gamma-Ray Space Telescope \cite{abdo2011fermi} in the 0.1-300 GeV energy range show a break in their spectra in the 1-10 GeV range. In their paper \cite{mena2013hints, mena2011signatures}, Mena, Razzaque, and Villaescusa-Navarro have modeled this spectral feature for the flat-spectrum radio quasar 3C454.3 during its November 2010 outburst, assuming that a significant fraction of the gamma rays convert to ALPs in the magnetic fields in and around the large scale jet of this blazar.

The main aim of this work is to follow the approach of \cite{mena2013hints} to explore the capability of an ALP-photon conversion model to produce the soft X-ray excess in the Coma cluster as claimed in \cite{angus2014soft}. This test is based upon whether this same model survives scrutiny in highly magnetized AGN jet environments. We find that, without accounting for misalignment, a model producing the Coma cluster X-ray excess over-produces X-rays in an environment like the M87 AGN. We aim to probe the effect of misalignment and off-axis jets in future work.

The structure of this paper is as follows. We briefly review the theoretical model that describes the photon-ALP mixing phenomena in Section \ref{sec.2} and discus whether this mixing can explain the soft X-ray excess phenomenon in Section \ref{sec.3}. The results are shown and discussed in Section \ref{sec.4}. Finally, a summary and conclusion are provided in Section \ref{sec.5}.

\section{ALPs-photon coupling model} \label{sec.2}

The coupling of an ALP with photon in the presence of an external magnetic field $\mathbf{B}$ is represented by the following Lagrangian \cite{sikivie1983experimental, raffelt1988mixing, anselm1988experimental}
\begin{equation}
\mathrm{\ell}_{a\gamma} = - \frac{1}{4} g_{a\gamma} \mathrm{F}_{\mu \nu} \tilde{\mathrm{F}}^{\mu \nu} a = g_{a\gamma}\, \mathbf{E} \cdot \mathbf{B} \, a \:,
\end{equation}
where $g_{a\gamma}$ is the the ALP-photon coupling constant, $\mathrm{F}_{\mu \nu}$ and $\tilde{\mathrm{F}}^{\mu \nu}$ represent the electromagnetic field tensor and its dual respectively, and $a$ donates the ALP field. While $\mathbf{E}$ and $\mathbf{B}$ are the electric and magnetic fields respectively. The evolution equations that describe the coupling of ALPs with a monochromatic and linearly polarized photon beam of energy $\omega$ propagating along the $z$-direction in the presence of an external and homogeneous magnetic field transverse $\mathbf{B}_T$ to the beam direction (i.e. in the x-y plane), takes the form \cite{raffelt1988mixing, bassan2010axion, mena2013hints, mena2011signatures}
\begin{equation} \label{eq.2}
i \dfrac{d}{dz} \left( \begin{matrix} A_{\perp}(z) \\ A{\parallel}(z) \\ a(z) \end{matrix} \right) = -  \left( \begin{matrix} 
\Delta_{\perp} \cos^2 \xi + \Delta_{\parallel} \sin^2 \xi & \cos \xi \sin \xi (\Delta_{\parallel}+\Delta_{\perp}) & \Delta_{a\gamma} \sin \xi  \\ 
\cos \xi \sin \xi (\Delta_{\parallel}+\Delta_{\perp})  & \Delta_{\perp} \sin^2 \xi + \Delta_{\parallel} \cos^2 \xi  & \Delta_{a\gamma} \cos \xi  \\
 \Delta_{a\gamma} \sin \xi & \Delta_{a\gamma} \cos \xi & \Delta_{a} \end{matrix} \right)
  \left( \begin{matrix} A_{\perp}(z) \\ A{\parallel}(z) \\ a(z) \end{matrix} \right) \: ,
\end{equation}
where $A_{\perp}$ and $A_{\parallel}$ are the photon linear polarization amplitudes along the $x$ and $y$ axis, respectively, and $a(z)$ donates the amplitude of ALPs. The parameter $\xi$ represents the angle between the transverse magnetic field $B_T$ and a fixed y-axis in the x-y plane.  For ALP-photon mixing model in the jet of the blazar 3C454.3, Mena and Razzaque adopted the following transverse magnetic field and electron density profiles \cite{mena2013hints}
\begin{equation}
\mathbf{B}_T = \phi \left( \frac{R}{10^{18} \text{cm}} \right)^{-1} \text{G} \:, \quad \text{and} \quad
n_e = \eta \left( \frac{R}{10^{18} \text{cm}} \right)^{-s} \text{cm}^{-3} \:,
\end{equation}
where $R$ is the radius from a central supermassive black hole, believed to be at the center of the AGN. The normalization parameters $\phi$ and $\eta$ are found in \cite{mena2013hints} by fitting GeV $\gamma$-ray data with this ALP-photon mixing model for different values of $s= 1,2$ and $3$ corresponding to different electron density profiles. The other different terms of the model, following Refs. \cite{bassan2010axion, mena2013hints, mena2011signatures}, are given as
\begin{align}
\Delta_{\perp} & \equiv 2 \Delta_{\text{QED}} + \Delta_{\text{pl}} \:, \qquad \Delta_{\parallel}  \equiv (7/2) \Delta_{\text{QED}} + \Delta_{\text{pl}} \:, \nonumber \\
\Delta_{QED}  &\simeq 1.34 \cdot 10^{-18} \phi^2 \left( \frac{\omega}{\text{GeV}} \right) \left( \frac{R}{10^{18} \text{cm}} \right)^{-2} \text{cm}^{-1} \:,  \nonumber \\
\Delta_{\text{pl}} &\simeq -3.49 \cdot 10^{-26} \eta \left( \frac{\omega}{\text{GeV}} \right)^{-1} \left( \frac{R}{10^{18} \text{cm}} \right)^{-2} \text{cm}^{-1} \:,  \nonumber \\
\Delta_{a\gamma} &\simeq 1.50 \cdot 10^{-17} \phi \left( \frac{g_{a\gamma}}{10^{-10}\text{GeV}^{-1}} \right) \left( \frac{R}{10^{18} \text{cm}} \right)^{-1} \text{cm}^{-1} \:,  \nonumber \\
\Delta_{a} &\simeq -2.53 \cdot 10^{-19} \left( \frac{\omega}{\text{GeV}} \right)^{-1} \left( \frac{m_a}{10^{-7} \text{eV}} \right)^{2} \text{cm}^{-1} \:.
\end{align}

Using this set of parameters, the evolution equations \ref{eq.2} can be numerically solved to find the two components of the photon linear politicization; $A_{\perp}$ and $A_{\parallel} $. Then the spectrum for a given blazar can be modified by a normalized suppression factor defined as
\begin{equation}
S(E) =  \vert A_{\parallel}(E) \vert^2 + \vert A_{\perp}(E) \vert^2  \:.
\end{equation}
Thus, the observation data be fitted using the ALP-photon mixing model by plotting the $\gamma$-ray energy spectra ($\nu F_{\nu} \equiv E^2 dN/dE$) as a function of the energy of the photons with $\omega \equiv E(1+z)$ where $z$ is the blazar's redshift. In the case of replicating \cite{mena2013hints} we note that the $\gamma$-ray energy spectra read
\begin{equation}
E^2 dN/dE = C E^{-\Gamma +2} S(E) \:.
\end{equation}
Here, $C$ and $\Gamma$ are spectral parameters. Hence, the ALP-photon mixing for the blazar jet model includes six free parameters: the normalization for the magnetic field $\phi$, the normalization for the electron density $\eta$, the ALP mass $m_a$, the ALP-photon coupling constant $g_{a\gamma}$, and two spectral parameters $C$ and $\Gamma$.

\section{Cosmic Axion Background} \label{sec.3}
In \cite{conlon2013excess}, the authors motivate the existence of a homogeneous Cosmic Axion Background (CAB) arising via the decay of string theory moduli in the very early universe. The suggestion being that the natural energy for such a background lies between $0.1$ and $1$ keV. Furthermore, in \cite{conlon2013cosmophenomenology} it was shown that such CAB would have a quasi-thermal energy spectrum with a peak dictated by the mass of the ALP. This CAB is also invoked in \cite{angus2014soft} to explain the soft X-ray excess on the periphery of the Coma cluster with an ALP mass of $1.1\times 10^{-13}$ eV and a coupling to the photon of $g = 2 \times 10^{-13}$ GeV$^{-1}$. 

In this work, we will assume, for convenience, that the CAB has a thermal spectrum with an average energy of $\langle E_a \rangle = 0.15$ keV. We then normalize the distribution to the typical example quoted in \cite{conlon2013cosmophenomenology}. We use the thermal distribution as an approximation, as the exact shape of the distribution will not substantially affect the conclusions we draw. We can then determine the fraction of CAB ALPs converted into photons within the environment of the M87 AGN jet and use this to determine a resulting photon flux. This flux can be compared to X-ray measurements to see if such environments can constrain low-mass ALPs and put limits on the ALP explanation of the Coma X-ray excess.

\section{Results and discussion} \label{sec.4}
The Fermi Large Area Telescope in the period from September 1st to December 13th, reported observations of the radio quasar 3C454.3 at redshift of z=0.895, constitutes of four epochs \cite{abdo2011fermi}: (i) A pre-flare period, (ii) A 13 day long plateau period, (iii) A 5 day flare, and (iv) A post-flare period. The ALP-photon mixing model then has been used to fit these observation data by plotting the $\gamma$-ray energy spectra ($\nu F_{\nu} \equiv E^2 dN/dE$) as a function of the photons energy $E$ in order to get some constraints on the ALP parameters. Among the whole analyses of the models, the photons have been considered to be initially unpolarized, and the following initial condition has been applied: $(A_{\parallel}(E), A_{\perp}(E), 0)=(\frac{1}{\sqrt{2}}, \frac{1}{\sqrt{2}}, 0)$ at $z \equiv R = 10^{18} \, \text{cm}$. In addition, the angel $\xi$ has been fixed to $\pi/4$ during the whole calculations. The evolution equations \ref{eq.2} have been solved numerically to fit with the spectral feature of the blazer 3C454.3 in its four epochs for three different cases of the electron profile density, as we explain downward here.

\begin{figure}[ht!]
\begin{minipage}{12.55pc}
\centering
\includegraphics[width=12.55pc]{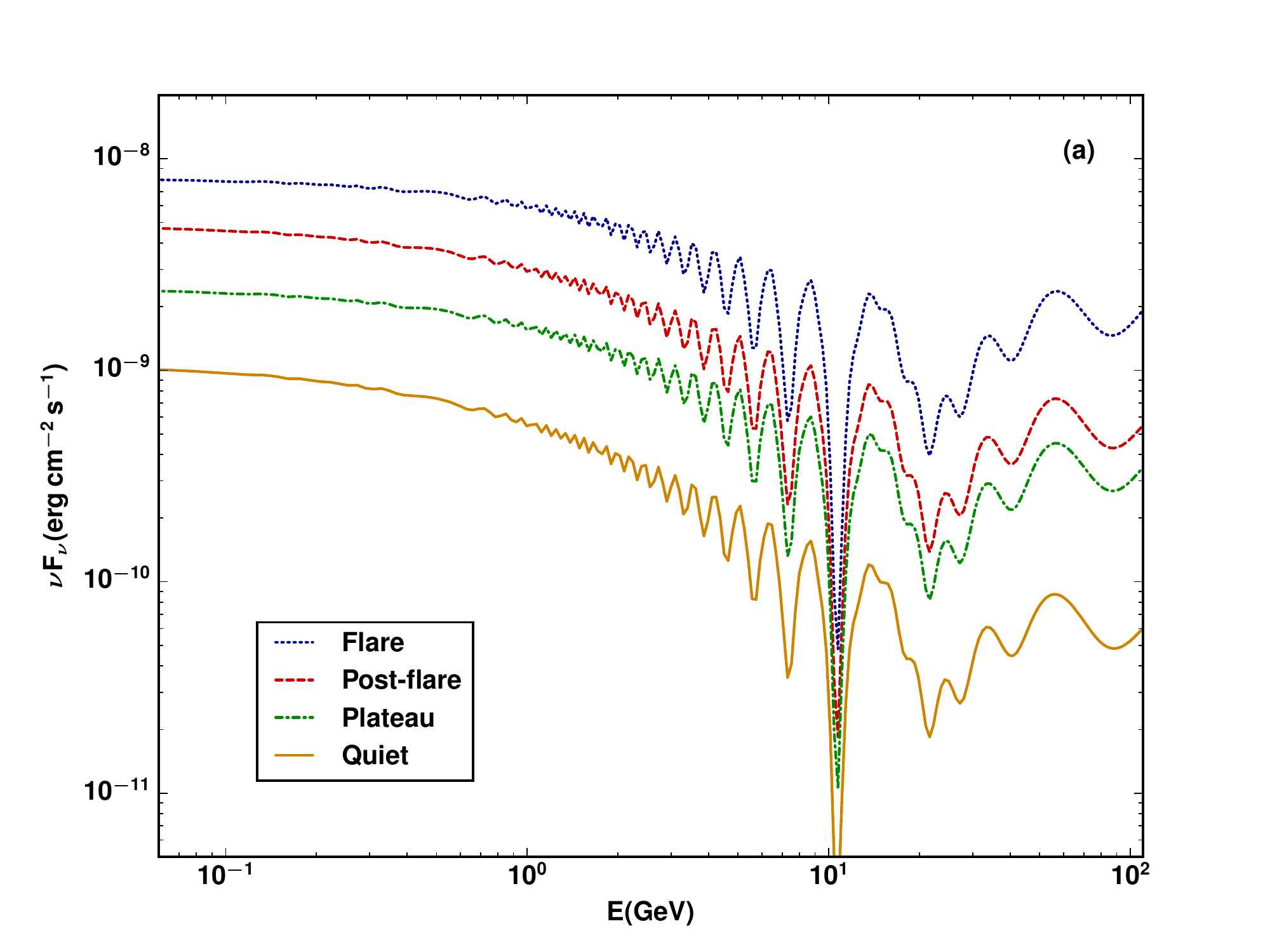}
\end{minipage}\hspace{0.2pc}
\begin{minipage}{12.55pc}
\centering
\includegraphics[width=12.55pc]{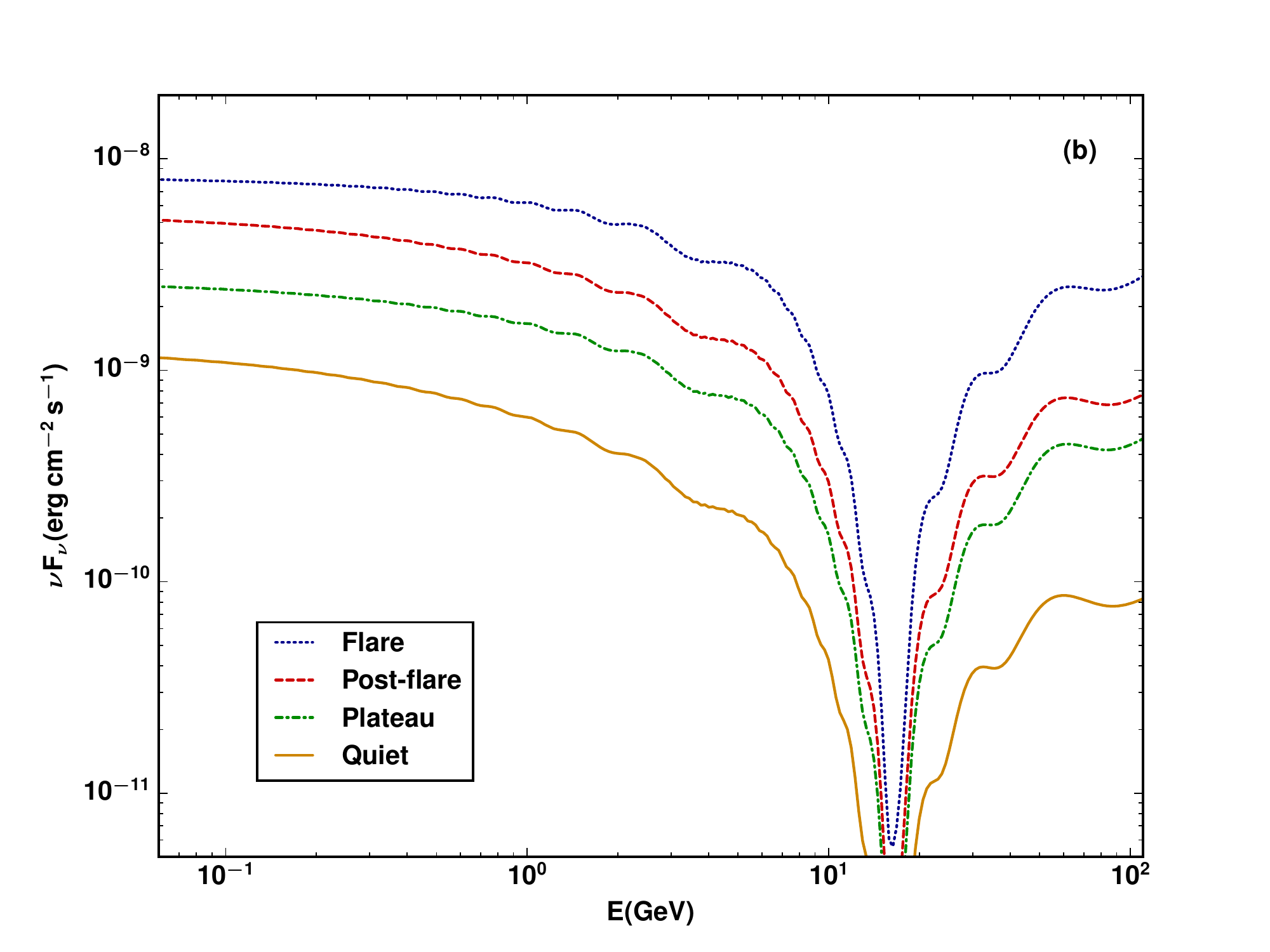} 
\end{minipage} \hspace{0.2pc}
\begin{minipage}{12.55pc}
\centering
\includegraphics[width=12.55pc]{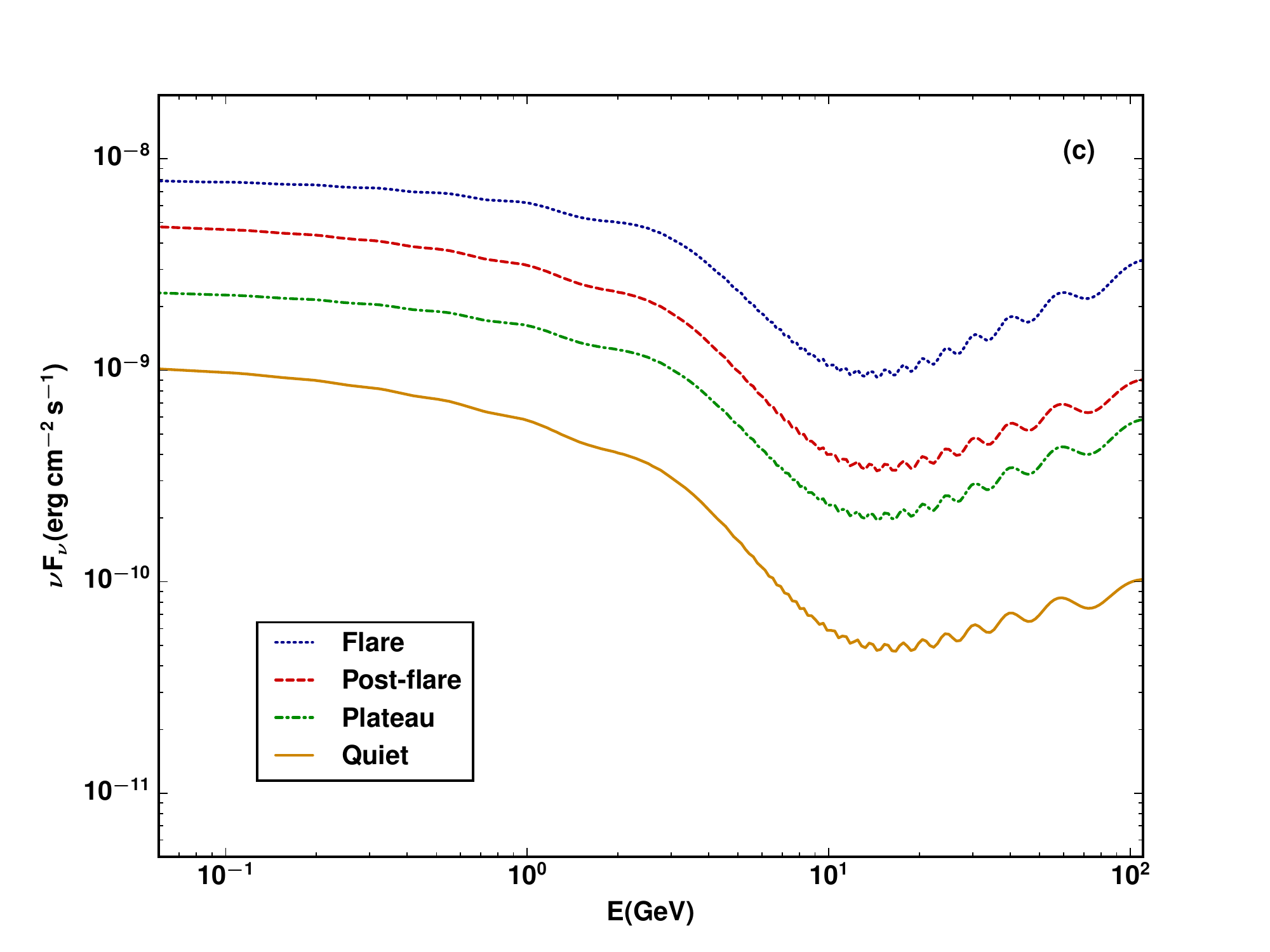}
\end{minipage} 
\caption{Fitting the spectral energy distributions to blazer 3C454.3 for four different epochs using ALP-photon mixing model with electron profile density: (a) $n_e \propto R^{-1}$, (b) $n_e \propto R^{-2}$  and (c) $n_e \propto R^{-3}$.}
\label{fig.1}
\end{figure}

Figures \ref{fig.1} show the fitting results of the spectral energy distributions to blazer 3C454.3 for four different epochs (flare, post-flare, plateau, and quiet) using the ALP-photon mixing model with the three different electron density profiles with $s=1$ (a), $2$ (b), and $3$ (c). The two spectral parameters $C$ and $\Gamma$ have been varied from epoch to another as they affected by the $\gamma$-ray emission region, while the other four parameters; $\phi$, $\eta$, $m_a$, and $g_{a\gamma}$ have been kept fixed. Thus, the best fitting of the ALP-photon mixing model for the observation data of blazar jet has been achieved  when the transition between photons to ALPs take place over different radii, $R \sim 10^{18}-10^{21} \, \text{cm}$ for $\phi \sim 10^{-2} \, \text{G}$, $\eta \sim 10^9 \, \text{cm}^{-3}$, $m_a \sim 10^{-7} \, \text{eV}$, and $g_{a\gamma}\sim 10^{-10} \, \text{GeV}^{-1}$. Comparing our results with the published results in Ref. \cite{mena2013hints}, allows us to deduce that the spectral energy distributions obtained using the numerical solutions to the evolution equations \ref{eq.2} show a good agreement with these published results which obtained based on fitting the observation data for the 3C454.3 blazar with the results of the ALP-photon mixing model.

\begin{figure}[ht!]
\centering
\includegraphics[scale=0.725]{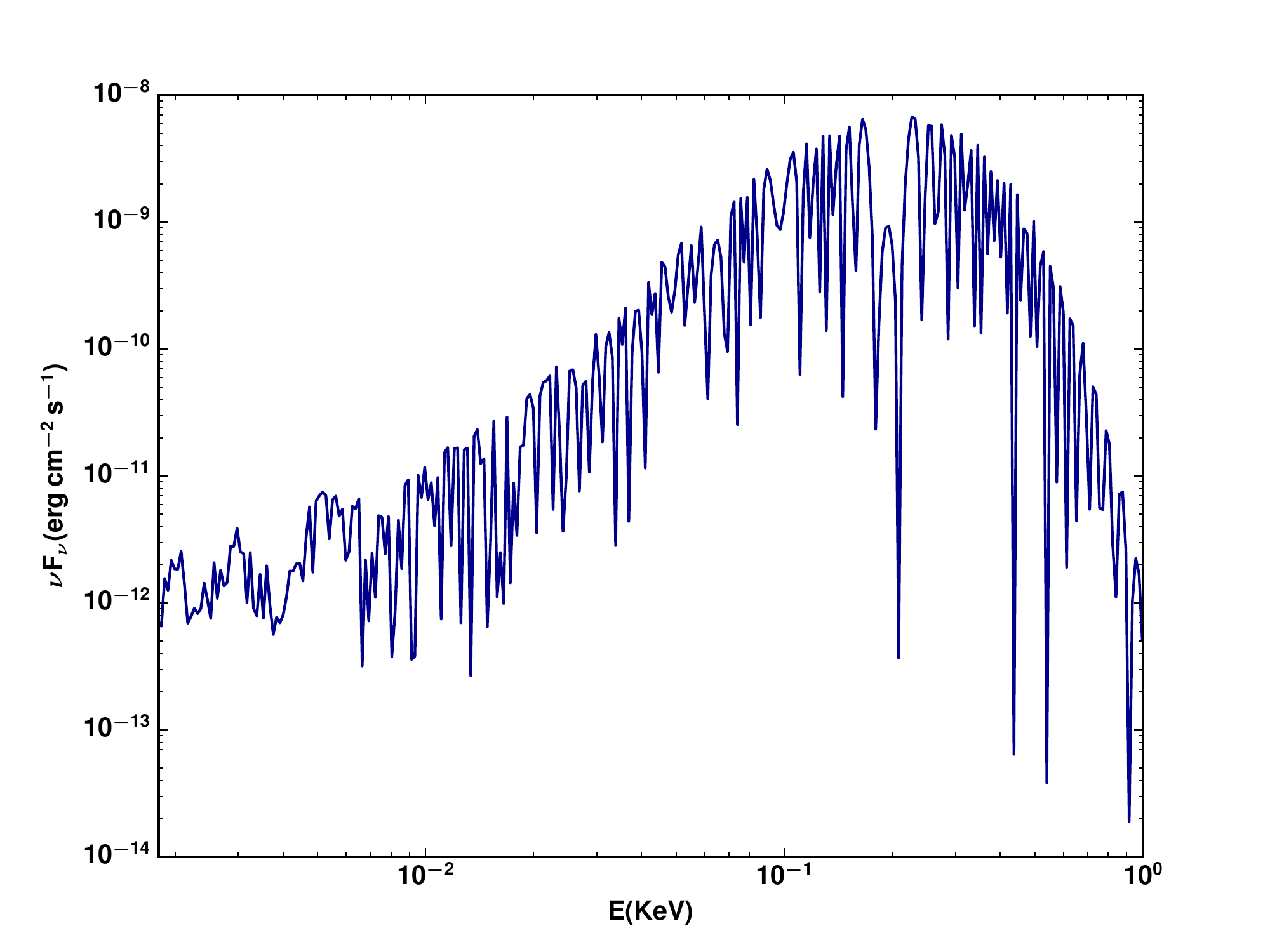}
\caption{The numerical simulation of the energy spectrum distributions from ALPs conversion to photons in the intergalactic magnetic field on the jet of M87 AGN.}
\label{fig.2}
\end{figure}
As a step forward, we apply the ALP-photon mixing model to study the probability of CAB ALPs to convert to photons in the intergalactic magnetic field on the jet of M87 AGN. We use an electron profile density profile of the first case with $s=1$. In addition, we set the ALP mass to be $1.1\times 10^{-13}$ eV and the ALP-photon coupling to be $g = 2 \times 10^{-13}$ GeV$^{-1}$ in agreement with the models derived in \cite{angus2014soft} to explain the soft X-ray excess on the periphery of the Coma cluster. The environmental parameters are taken as $\phi= 1.4 \times 10^{-3}$ and $\eta=0.3$ to make sure that the magnetic field and the electron density profiles used for M87 AGN are consisted with the obtained values in \cite{park2019faraday}. Preliminary results are showing in Fig. \ref{eq.2} for the energy spectrum distributions obtained from the numerical simulation for the ALPs conversion to photons in the intergalactic magnetic field of the jet of M87 AGN. It is evident in these results that we over-produce the observed AGN flux in M87, between 0.3 and 8 keV,  by two orders of magnitude~\cite{m87chandra}. 

\section{Conclusion} \label{sec.5}

In this work we have examined the ALP-photon model which developed by Mena and Razzaque to fit the spectral feature for the  the flat-spectrum radio quasar 3C454.3 during its November 2010 outburst in order to get some constraints on the ALP parameters by assuming that a significant fraction of the gamma rays convert to ALPs in the presence of external magnetic field in the large scale jet of this blazar. We successfully reproduced their results with a very good agreement with the observation data for the 3C454.3 blazar. Indeed, this makes us very confident that the simulation is robust. As a step forward, we used the model to test whether the CAB ALP conversion to photons can generate the soft X-ray excess in the environment of the Coma galaxy cluster. We find that the environment of the M87 AGN jet provides preliminary suggestions of an X-ray over-production around two orders of magnitude. Further effects, such as misalignment, must be considered before we can rule out the ALP model proposed to account for the observed Coma X-ray excess.

\section{Acknowledgments}

This work is based on the research supported by the South African Research Chairs Initiative of the Department of Science and Technology and National Research Foundation of South Africa (Grant No 77948). A. Ayad acknowledges support from the Department of Science and Technology/National Research Foundation (DST/NRF) Square Kilometre Array (SKA) post-graduate bursary initiative under the same Grant. G. Beck acknowledges support from a National Research Foundation of South Africa Thuthuka grant no. 117969. The authors would like also to offer special thanks to Prof. S. Colafrancesco, who, although no longer with us, continues to inspire by his example and dedication to the students he served over the course of his career. 

\section*{References}

\bibliography{references}

\providecommand{\newblock}{}
\begin{thebibliography}{10}
\expandafter\ifx\csname url\endcsname\relax
  \def\url#1{{\tt #1}}\fi
\expandafter\ifx\csname urlprefix\endcsname\relax\def\urlprefix{URL }\fi
\providecommand{\eprint}[2][]{\url{#2}}
% Bibliography created with iopart-num v2.0
% /biblio/bibtex/contrib/iopart-num

\bibitem{komatsu2011seven}
Komatsu E, Smith K, Dunkley J, Bennett C, Gold B, Hinshaw G, Jarosik N, Larson
  D, Nolta M, Page L {\em et~al.\/} 2011 {\em The Astrophysical Journal
  Supplement Series\/} {\bf 192} 18

\bibitem{peccei1977cp}
Peccei R~D and Quinn H~R 1977 {\em Phys. Rev. Lett.\/} {\bf 38} 1440--1443

\bibitem{weinberg1978new}
Weinberg S 1978 {\em Physical Review Letters\/} {\bf 40} 223

\bibitem{peccei2008strong}
Peccei R~D 2008 {\em Axions\/} (Springer) pp 3--17

\bibitem{asztalos2006searches}
Asztalos S~J, Rosenberg L~J, van Bibber K, Sikivie P and Zioutas K 2006 {\em
  Annu. Rev. Nucl. Part. Sci.\/} {\bf 56} 293--326

\bibitem{preskill1983cosmology}
Preskill J, Wise M~B and Wilczek F 1983 {\em Physics Letters B\/} {\bf 120}
  127--132

\bibitem{abbott1983cosmological}
Abbott L~F and Sikivie P 1983 {\em Physics Letters B\/} {\bf 120} 133--136

\bibitem{dine1983not}
Dine M and Fischler W 1983 {\em Physics Letters B\/} {\bf 120} 137--141

\bibitem{arvanitaki2010string}
Arvanitaki A, Dimopoulos S, Dubovsky S, Kaloper N and March-Russell J 2010 {\em
  Physical Review D\/} {\bf 81} 123530

\bibitem{cicoli2012type}
Cicoli M, Goodsell M~D and Ringwald A 2012 {\em Journal of High Energy
  Physics\/} {\bf 2012} 146

\bibitem{anselm1982second}
Anselm A and Uraltsev N 1982 {\em Physics Letters B\/} {\bf 114} 39--41

\bibitem{arias2012wispy}
Arias P, Cadamuro D, Goodsell M, Jaeckel J, Redondo J and Ringwald A 2012 {\em
  Journal of Cosmology and Astroparticle Physics\/} {\bf 2012} 013

\bibitem{ringwald2012exploring}
Ringwald A 2012 {\em Physics of the Dark Universe\/} {\bf 1} 116--135

\bibitem{marsh2016axion}
Marsh D~J 2016 {\em Physics Reports\/} {\bf 643} 1--79

\bibitem{sikivie1983experimental}
Sikivie P 1983 {\em Physical Review Letters\/} {\bf 51} 1415

\bibitem{raffelt1988mixing}
Raffelt G and Stodolsky L 1988 {\em Physical Review D\/} {\bf 37} 1237

\bibitem{bassan2010axion}
Bassan N, Mirizzi A and Roncadelli M 2010 {\em Journal of Cosmology and
  Astroparticle Physics\/} {\bf 2010} 010

\bibitem{horns2012probing}
Horns D, Maccione L, Mirizzi A and Roncadelli M 2012 {\em Physical Review D\/}
  {\bf 85} 085021

\bibitem{harris2014photon}
Harris J and Chadwick P~M 2014 {\em Journal of Cosmology and Astroparticle
  Physics\/} {\bf 2014} 018

\bibitem{abdo2011fermi}
Abdo A~A, Ackermann M, Ajello M, Allafort A, Baldini L, Ballet J, Barbiellini
  G, Bastieri D, Bellazzini R, Berenji B {\em et~al.\/} 2011 {\em The
  Astrophysical journal letters\/} {\bf 733} L26

\bibitem{mena2013hints}
Mena O and Razzaque S 2013 {\em Journal of Cosmology and Astroparticle
  Physics\/} {\bf 2013} 023

\bibitem{mena2011signatures}
Mena O, Razzaque S and Villaescusa-Navarro F 2011 {\em Journal of Cosmology and
  Astroparticle Physics\/} {\bf 2011} 030

\bibitem{angus2014soft}
Angus S, Conlon J~P, Marsh M~D, Powell A~J and Witkowski L~T 2014 {\em Journal
  of Cosmology and Astroparticle Physics\/} {\bf 2014} 026

\bibitem{anselm1988experimental}
Anselm A~A 1988 {\em Physical Review D\/} {\bf 37} 2001

\bibitem{conlon2013excess}
Conlon J~P and Marsh M~D 2013 {\em Physical review letters\/} {\bf 111} 151301

\bibitem{conlon2013cosmophenomenology}
Conlon J~P and Marsh M~D 2013 {\em Journal of High Energy Physics\/} {\bf 2013}
  214

\bibitem{park2019faraday}
Park J, Hada K, Kino M, Nakamura M, Ro H and Trippe S 2019 {\em The
  Astrophysical Journal\/} {\bf 871} 257

\bibitem{m87chandra}
{Donato} D, {Sambruna} R~M and {Gliozzi} M 2004 {\em ApJ\/} {\bf 617} 915--929
  (\textit{Preprint} \eprint{astro-ph/0408451})

\end{thebibliography}

\end{document}